\documentclass{article}

\usepackage{arxiv}

\usepackage[utf8]{inputenc} 
\usepackage[T1]{fontenc}    
\usepackage{hyperref}       
\usepackage{url}            
\usepackage{booktabs}       
\usepackage{amsfonts}       
\usepackage{nicefrac}       
\usepackage{microtype}      
\usepackage{lipsum}		
\usepackage{graphicx}
\usepackage{natbib}
\usepackage{doi}
\usepackage[ruled,vlined]{algorithm2e}

\usepackage{amssymb}
\usepackage{amsmath}
\usepackage{mathrsfs}
\usepackage{amsfonts}
\usepackage{xspace}
\usepackage{systeme}
\usepackage{setspace}
\usepackage{tabularx}

\title{A deep learning pipeline for breast cancer Ki-67 proliferation index scoring}

\author{Khaled Benaggoune \thanks{Coresponding author: k.benaggoune@univ-batna2.dz.} \\
	FEMTO-ST institute,\\
	Univ. Bourgogne Franche-Comte,\\
	CNRS, ENSMM, Besan\c{c}on, France\\
	\texttt{k.benaggoune@univ-batna2.dz,} \\
	\texttt{khaled.benaggoune@femto-st.fr}\\
	\And
	Zeina Al Masry \\
	FEMTO-ST institute,\\
	Univ. Bourgogne Franche-Comte,\\
	CNRS, ENSMM, Besan\c{c}on, France\\
	\texttt{zeina.almasry@femto-st.fr} \\
	\And
	Jian Ma \\
	School of Reliability and Systems Engineering,\\
	Beihang University, Beijing, China,\\
	\texttt{majian3128@126.com} \\
	\And
	Christine Devalland \\
	Service D'anatomie Et Cytologie Pathologiques,\\
	H\^opital Nord Franche-Comt\'e, Trevenans, France\\
	\texttt{christine.devalland@hnfc.fr} \\
	\And
	Leila Hayet Mouss\\
	Laboratory of Automation and Production engineering,\\
	Batna2 University,Batna, Algeria\\
	\texttt{hayet\_mouss@yahoo.fr}\\
	\And
	Noureddine Zerhouni\\
	Univ. Bourgogne Franche-Comte,\\
	CNRS, ENSMM, Besan\c{c}on, France\\
	\texttt{noureddine.zerhouni@femto-st.fr}\\
}

\renewcommand{\shorttitle}{A deep learning pipeline for breast cancer Ki-67 proliferation index scoring}

\hypersetup{
pdftitle={A pipeline for breast cancer Ki-67 proliferation index scoring using deep learning techniques},
pdfsubject={cs.CV},
pdfauthor={Khaled Benaggoune, Zeina Al Masry, Christine Devalland, Leila HayetMouss, Noureddine Zerhouni},
pdfkeywords={Deep learning, Cell counting, Ki-67 index, Breast cance, Segmentation, Classification.},
}

\begin{document}
\maketitle

\begin{abstract}
The Ki-67 proliferation index is an essential biomarker that helps pathologists to diagnose and select appropriate treatments. However, automatic evaluation of Ki-67 is difficult due to nuclei overlapping and complex variations in their properties. This paper proposes an integrated pipeline for accurate automatic counting of Ki-67, where the impact of nuclei separation techniques is highlighted.  First, semantic segmentation is performed by combining the Squeez and Excitation Resnet and Unet algorithms to extract nuclei from the background. The extracted nuclei are then divided into overlapped and non-overlapped regions based on eight geometric and statistical features. A marker-based Watershed algorithm is subsequently proposed and applied only to the overlapped regions to separate nuclei. Finally, deep features are extracted from each nucleus patch using Resnet18 and classified into positive or negative by a random forest classifier. The proposed pipeline's performance is validated on a dataset from the Department of Pathology at Hôpital Nord Franche-Comté hospital.

\end{abstract}


\section{Introduction}
\label{S:1}
Breast cancer is the most common cancer with high mortality among women worldwide \citep{stewart2014world}. Early detection of this disease leads to improve results and to extend the survival rate \citep{jemal2009cancer}. Ki-67 is a nuclear antigen linked to the cell cycle that measures cellular proliferation rate, usually evaluated by using Immunochemistry. However, there are several methods to estimate this index,which differ on how regions are selected, the number of regions, and nuclei labeling affected by stain intensity variation. In \citep{besusparis2016impact}, the authors demonstrated on 297 breast cancer sections that tumors with low proliferative activity would require a larger sampling size to obtain the same error-measurement results as for highly proliferative tumors. Therefore, the optimal tissue sampling for the evaluation of IHC biomarkers depends on the studied tissue's heterogeneity and must be determined on a use-by-use basis. In spite of numerous data on  Ki-67 as a prognostic marker in breast cancer, the inter-variability and intra-variability in the manual counting limit the accuracy of this scoring and its application in the treatment. Accordingly, the Ki-67 Proliferation Index (PI) is calculated by averaging the percentage of immunopositive nuclei to the total nuclei in all selected regions. We can recognize different methods proposed in the literature, such as including counting Ki-67 immunopositive cells in 10 high power fields \citep{goodell2012comparison}, eyeballing (i.e., best estimate) \citep{phan2010nanets}, counting 2000 cells in regions of interest (i.e. hotspots \citep{khan2014perceptual}) with the most frequent Ki-67 nuclear labeling \citep{scholzen2000ki}, or counting using automated image analysis \citep{lopez2012clustering}. In the last decade, digital pathology (DP) is becoming increasingly common due to the growing availability of whole slide digital scanners \citep{ghaznavi2013digital}, and it could be a good candidate to resolve the related issues of variability and time optimization for PI computation. It simplifies the storage and sharing of stained tumor slides, and make the automatic analysis of very large cohorts possible. 

Recently, deep learning has been applied widely in the medical field \citep{zemouri2018constructive, zuluaga2019survey, ma2019portable}, especially in automatic analysis of histopathological images \citep{abubakar2016high}. Many studies have noted that automating the process of PI analysis yield more reproducible and accurate measurement \citep{harvey2015practical, gudlaugsson2012comparison}, and its application is being implemented in a clinical setting. In breast cancer cell microscopy, many works such as segmentation, detection, and classification of different types of cancerous cells are proposed to support and automate diagnosis cancer systems. As for the histopathological images, segmentation and classification of cells still challenging and require more development due to the random variations in the morphology, the color, and the intensity of cells. {In some practical cases, these cells occupy a small part of the image, yielding some problems such as cell clusters and a class imbalance between the foreground and background. Hence, segmentation of the overlapped regions is a major problem in cell counting.} To be more precise, counting several overlapped cells as one cell may affect the accuracy of the PI which can negatively influence the pathologist's decisions. Many works have already studied this type of indicator in order to automate this score \citep{geread2019ihc, shi2016automated, barricelli2019ki67}. Nevertheless, there are more issues to solve to get more accurate results, such as cells' superposition, color variation, etc. This study highlights the impact of overlapped nuclei separation on the PI scoring by introducing post-processing between nuclei detection and nuclei classification steps. 

In this paper, we are interested in providing an accurate PI score by considering the overlapped nuclei issue. An automatic approach is proposed using a new private dataset. First regions of interest are manually selected by an expert pathologist. Then, for nuclei segmentation, a U-shape fully convolutional network called U-net \citep{ronneberger2015u} is used with squeeze and excitation resnet backbone  SER-Unet to separate nuclei from the background. The predicted mask is processed in order to separate overlapped nuclei regions from single nuclei regions. After that, a marker-based watershed method is proposed to split overlapped nuclei. In the last step, Resnet18 is used to extract deep features from nuclei patches, and rendom forest classifier is then used to assign each nucleus into immunopositive or immunonegative in order to calculate the PI score. 

The remainder of this paper is as follows. Section \ref{RW} presents the related work as well as the position contribution. In Section \ref{approach}, we provide a description of the proposed approach to compute the PI scoring. A case study is then presented in Section \ref{results}. Finally, a discussion and a conclusion are detailed in Section \ref{disc}.

\section{Related Work}
\label{RW}
Recently, the emergence of DP and the availability of a large number of digitized scans have initiated a flow of new methodological developments in computer-aided diagnosis \citep{doi2007computer}. Popular topics in this field include image pre-processing, cell detection \citep{irshad2013methods}, classification of tissue regions, or prediction of tumor types \citep{hayakawa2019computational}. 

\begin{figure}[h]
\centering
        \includegraphics[totalheight=5cm]{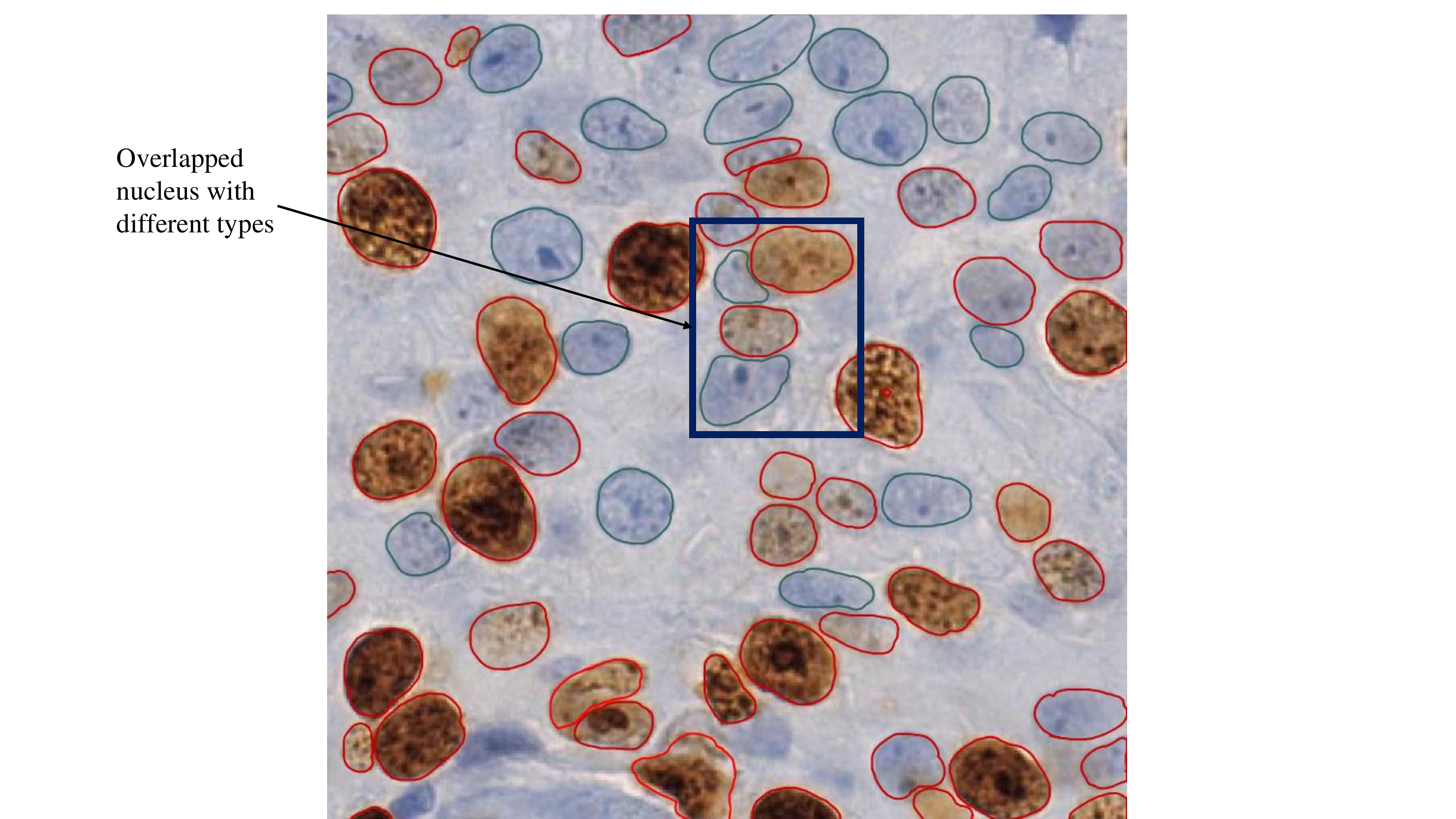}
    \caption{Overlapped nuclei from different types. immunopositive nuclei are annotated with red and immunonegative nuclei with green.}
    \label{fig:overlapped}
\end{figure}

The segmentation of nuclei is an essential first step towards the automatic analysis of PI. It is a challenging phase in the field of DP for numerous reasons: First, the morphology of the nuclei is a critical element of most cancer classification systems. Second, effective nucleus segmentation techniques can efficiently diminish the human effort required for cell-level analysis.  Therefore, the cost of cellular analysis can be significantly optimized. However, the segmentation of nuclei presents many difficulties, such as the identification of overlapping nuclei and finding a precise delineation framework \citep{caicedo2019nucleus}. A straightforward way to avoid the problem of overlapped cell separation is to estimate the number of cells in each detected region without splitting. However, this can mislead the classification of the block of cells including overlapped nucleus from different types (Figure \ref{fig:overlapped}), and incorrect segmentation and detection of nuclei lead to a wrong PI estimation. 

{Recently, deep learning, especially convolutional neural network (CNN), has improved medical image analysis performance.} A survey of deep learning in image cytometry is presented in \citep{gupta2019deep}. Counting cells in histopathology images can be viewed as giving the object localization without accurately delineating boundaries, usually known as a marker or cell centroid. In \citep{xie2018microscopy}, the authors addressed the cell counting task as a regression problem, where they used a fully convolutional network (FCN) to regress a cell spatial density map across the image. The FCN is trained in synthetic data but still gives good results without tuning in real data. Counting cells by regression may be useful in cases of overlapped cells, but it does not give relevant information about the cell morphology, which is necessary to the next step of cell classification in PI estimation. Hence, for better PI estimation, we need to separate cells accurately from the background. First attempts to segment cells using pixel sliding window CNN suffer from the contextual information sacrifice for location accuracy in small patches  \citep{gupta2019deep}. To overcome this issue, authors in \citep{long2015fully} presented a FCN trained end to end, pixels by pixels on semantic segmentation. This architecture with successive convolutional layers increases the resolution of the output. \citep{ronneberger2015u} proposed a new network architecture called U-net that can work with very few training images and yields more precise segmentation. Subsequently, numerous architectures based on FCN are proposed for segmentation tasks like SegNet \citep{badrinarayanan2017segnet}, LinkNet \citep{chaurasia2017linknet}, Tiramisu \citep{jegou2017one}, DilatedNet \citep{yu2015multi}, PSPNet \citep{zhao2017pyramid}. However, even with this advance in deep learning segmentation architectures, the overlapped cell delineation still an ongoing problem \citep{moen2019deep}. In most cases, the predicted masks are post-processed with different techniques to separate cells. Simple filters like erosion and morphological opening and closing could help in removing artifacts, but cannot separate overlapped cells in most cases.

Extensive and targeted research has been carried out by various groups on nuclear cluster splitting \citep{irshad2013methods}. Despite the fact that it is not always possible to solve the overlap problem, such splits are often essential to extract additional features useful for the next step of nuclei classification, even when they come from segmentations that include some mis-assigned pixels. Proposed approaches either use the cluster's contour concavity measures to identify nuclei or apply ellipse fitting to infer overlapping nuclei. One drawback of these methods is that performance tends to be affected by contour fluctuations \citep{plissiti2014splitting, zafari2015segmentation}. The watershed algorithm is one of the popular techniques used to separate clusters of nuclei \citep{beucher1979use}.  Watershed deals with the separation of regions by creating unique containers where nuclei are separated by the edges that delineate the containers. While this can enable successful separation in some images, there are well-documented issues of over- and under-segmentation in nuclei overlapping, mainly when the boundary gradients at the overlaps are small \citep{ali2011medical}. To overcome this problem, \citep{wahlby2004combining} used morphological filtering to find the seeds of nuclei. Then, segmentation of the seeded watersheds is applied to the gradient magnitude image to create the region's boundaries. Later, the initial segmentation result is refined with the gradient magnitude along the edge separating neighboring objects, removing poorly contrasted objects. Finally, the distance transformation and shape-based cluster separation methods are applied in the final step, keeping only the separation lines, which have crossed the distance map's deep valleys.  \citep{cloppet2008segmentation} presented a segmentation scheme of overlapping nuclei in immunofluorescence images by giving a specific set of markers to the watershed algorithm to split the overlapping structures. 

After nuclei are precisely segmented, classification is the second step towards the automatic analysis of PI. Features computed from segmented nuclei are usually required for nuclei classification that make higher-level information about the disease state. The classifiers use nuclei features, which indicate the deviations in the nuclei structures, to learn how to classify nuclei into different classes. In histopathology, these features can be categorized into the following four categories: 1) cytological; 2) intensity; 3) morphological; and 4) texture features \citep{irshad2013methods}. In \citep{huang2010effective}, the total of 14 features of intensity, morphological, and texture are extracted to train a Support Vector Machine (SVM) based graph classifier and compared with K-nearest neighbor and simple SVM.  In \citep{geread2019ihc}, an unsupervised color separation method based on the IHC color histogram (IHCCH) is proposed for the nuclei segmentation and classification, with robust analysis of Ki67 datasets. An overstaining threshold is implemented to adjust for background overstaining. The method accurately quantified the PI over the dataset, with an average proliferation index difference of 3.25\%. In \citep{shi2016automated} The classification strategy mainly consists of smoothing, decomposition, feature extraction, and K-means clustering. \citep{irshad2013multi} used intensity and morphology features with a decision tree and SVM classifiers for mitosis detection in breast cancer histopathology images. The first attempt to use deep learning in PI estimation was in \citep{saha2017advanced}. The authors used the Gamma mixture model with Expectation-Maximization for seed point detection and patch selection. {For classification, a deep learning model is used, and for proliferation scoring, a decision tree is used as a final decision layer.} In \citep{narayanan2018deepsdcs} the same deep learning model is used for nuclei detection and classification. VGG16 layers are used to extract hypercolumn descriptors from each cell to form the vector of activation. Subsequently, seeds made from cell segmentation were propagated to a spatially constrained convolutional neural network for cell classification.

In general, deep learning methods give better results in nuclei segmentation and classification. However, using one model for both segmentation and classification is not suitable for PI estimation due to the complexity of the two tasks and overlapped nuclei. Moreover, most of the proposed deep learning models in PI estimation are used for one task, segmentation or classification, combined with other conventional techniques. In the sequel, we propose a new approach using CNNs models for nuclei segmentation and classification. CNN methods overcome the problem of color variation, stain intensity and perform feature extraction automatically. However,  since the semantic segmentation results affect the PI scoring,  an intermediate step is proposed between the two models to enhance the segmentation; therefore, the classification and the counting results.

\section{Materials and methods}
\label{approach}

{\subsection{Materials}
The dataset is composed of tumors formalin-fixed paraffin-embedded tissues from invasive breast carcinoma from HNFC hospital. 4 $\mu$m sections were cut and used for immunochemistry. Theses excision slides  were then processed for immunochemistry with benchmark Ventana automate. The monoclonal antibody KI67 (Dako clone Mib one) was applied at a 1/50 dilution for 20 minutes. 
}

{Fifty slides were captured at 40x magnification on a Hamamatsu scanner. Three sub-images of size 256x256 pixels were cropped from three regions of interest Hot-spot (the area of higher density of positive tumor cells for Ki-67), edges, and medium (areas of medium density of positive tumor cells for Ki-67) selected from each patient to ensure the data variation (see \citep{jang2017comparison}). After obtaining 150 from all cases, an expert in pathology, a resident student, and a biomedical engineer have annotated more than 5000 nuclei. The open-source software Qupath is used for annotation using the Brush tool on 25" monitor \citep{bankhead2017qupath}. The annotators are asked to delineate all nucleus' boundaries and associate each nucleus to its class positive or negative. The final annotation of the resident student and the biomedical engineer is revised by the pathologist expert to reduce false positive, false negative, under-segmented, and over-segmented errors. 
Two datasets are generated using Qupath tool from the annotated data:
\begin{itemize}
    \item 150 binary masks are extracted from the annotated datasets to train the segmentation model in a supervised manner. For models training and validation, the dataset is split into 30, 10, and 10 patients, with three images for each patient, for training, validation, and test sets, respectively.
    \item For classification, using the Opencv library \citep{howse2013opencv}, we extract immunopositive and immunonegative nuclei patches with size 32x32, which can include the whole nuclei at 40 magnification. The total of 5167 patches is divided by patient to avoid overfitting to 3214, 931, 1022 for train, validation and test.
\end{itemize}}
\subsection{Methods}
\begin{figure}[h]
\centering
        \includegraphics[totalheight=14cm]{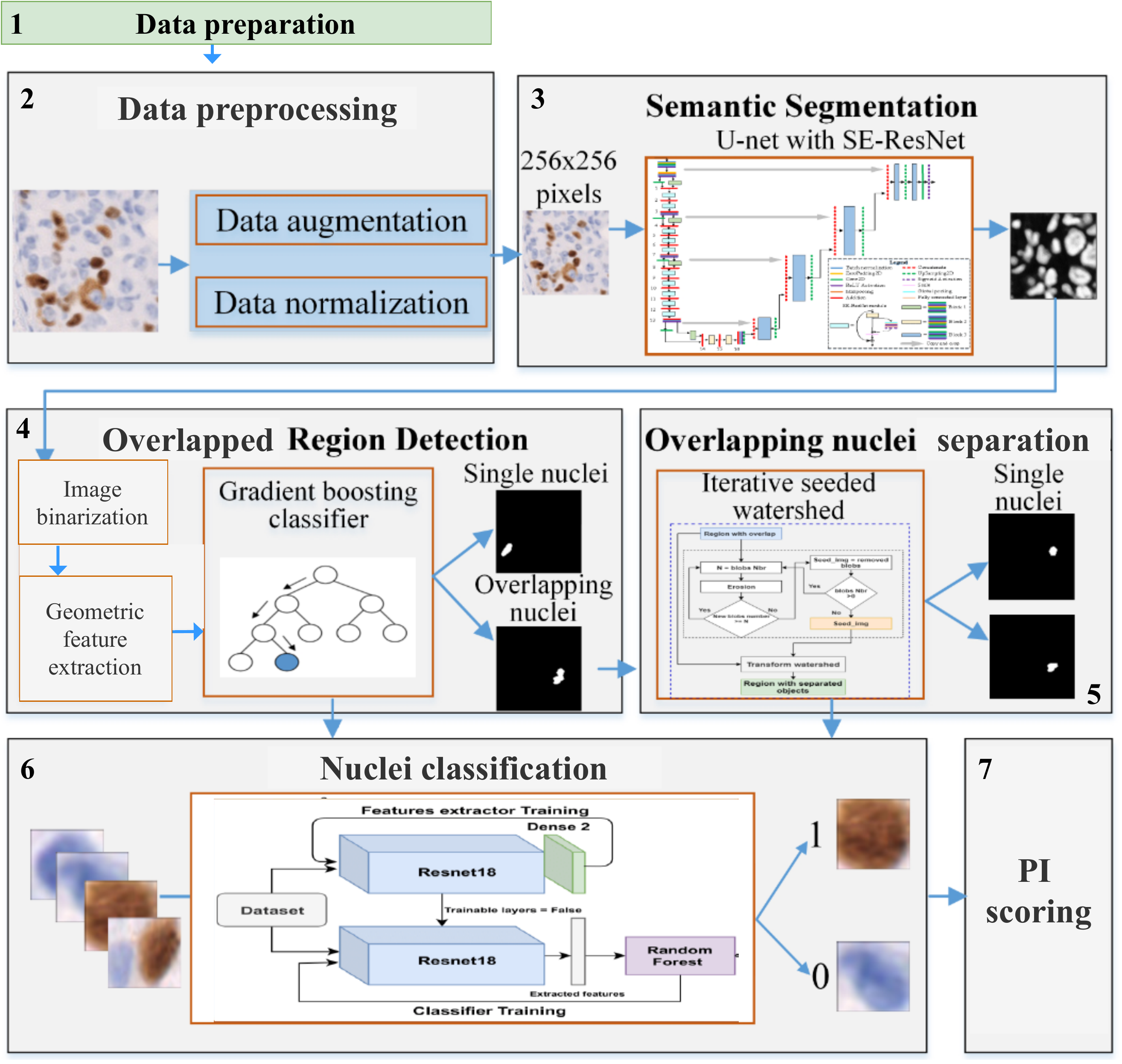}
    \caption{Flowchart of the proposed cell counting methods from data preprocessing to PI scoring.}
    \label{fig:nn}
\end{figure}

To deal with stain and shape variability of nuclei more efficiently, a series of automated processes
are illustrated in Figure \ref{fig:nn}. First, data augmentation and normalization techniques are used to diminish the images' differences and enlarge the data volume. For the segmentation task,  SER-Unet is used to separate nuclei regions from the background. The segmented nuclei regions are classified by a gradient boosting classifier using extracted eight geometric features to be single-nucleus regions or overlapped nuclei regions. Considering the overlapped nuclei regions have a significant effect on the PI, a marker-based watershed model is used to separate overlapped nuclei. After that, all the single nuclei are identified by a CNN model to be positive or negative. Finally, the PI is calculated.

\subsubsection{Data preprocessing}

\begin{figure}[h]
\centering
        \includegraphics[totalheight=6cm]{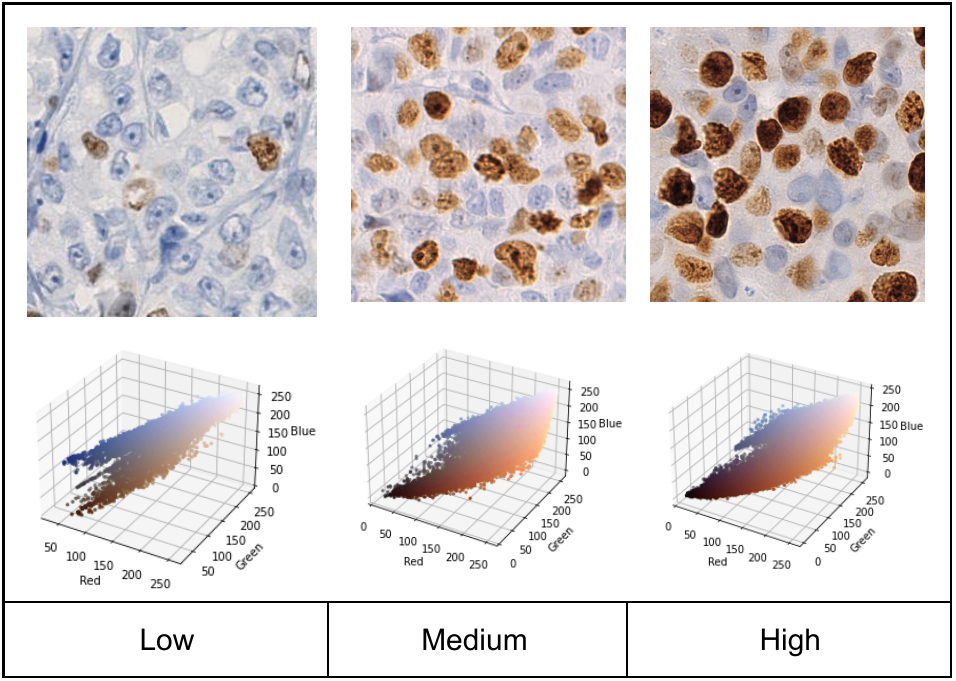}
    \caption{Stain variation in H-DAB images.}
    \label{stain_dif}
\end{figure}

The different use of scanners, equipment, and different stain coloring and reactivity generate color variation in histopathology images (Figure \ref{stain_dif}).  Color variation can unfavorably affect the training and inference process in machine learning models. Therefore, it is necessary to improve image quality before the automatic feature extraction. Data is normalized to eliminate the differences in the histology images' color. Color normalization techniques give a more general form of color correction that borrows one image's color characteristics from another. This paper adopts the image processing method called Structure preserving color normalization \citep{vahadane2016structure}. This method uses a simple statistical analysis and can achieve color correction by choosing a suitable source image and apply its characteristics to another image.

Besides, and since we have limited data, data augmentation is performed to train models for discriminative information learning. Data augmentation is done on each image using Albumentations library \citep{2018arXiv180906839B} to achieve the transformation in a pipeline fashion, where each operation has a probability that defines the likelihood of being applied. Two types of data augmentation are adopted: (i)- for classification task: simple augmentations such as random rotation, flip, transpose, and shift are used. (ii)- for segmentation task: strong augmentations are performed with Gaussian noise, blur, optical distortion, grid distortion, and elastic transformation. It is worth mentioning that data augmentation is carried out before data normalization in our pipeline. Finally, image pixel intensity is normalized to the standard normal distribution of the range of [0,1] to obtain the same range of values for each input image, which speeds up the model's convergence. 

\subsubsection{Semantic segmentation}

\begin{figure}[h]
\centering
        \includegraphics[totalheight=10cm]{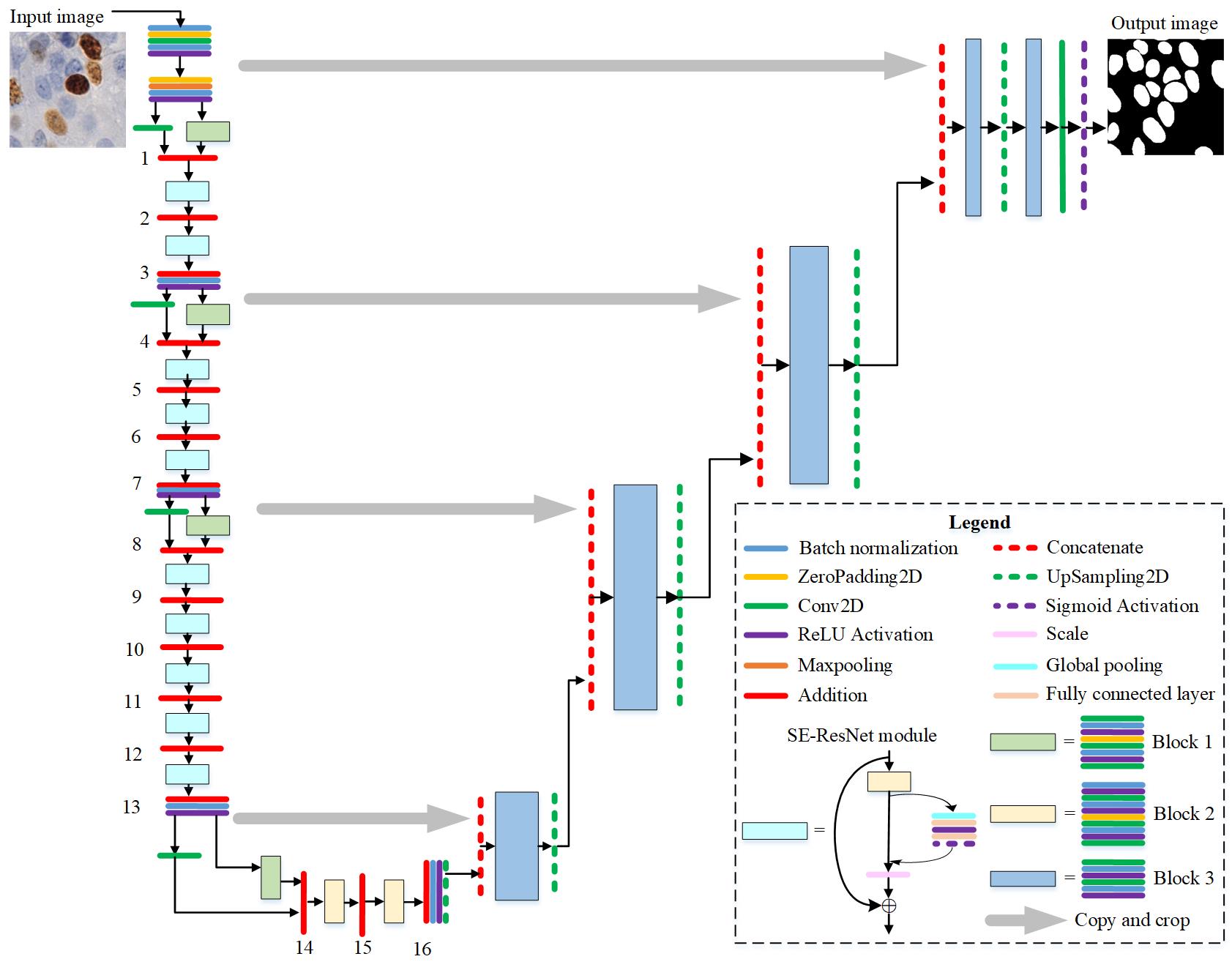}
    \caption{Semantic segmentation using U-net with Squeez and Excitation ResNet backbone.}
    \label{unet}
\end{figure}

 SER-Unet is here used to segment nuclei in histopathology images. U-net is the improvement and extension of FCNs \citep{long2015fully} with a more elegant architecture. It works with very few training images and yields more precise segments \citep{ronneberger2015u}. U-net's general architecture consists of two paths, a contracting path (left side) and an expansive path (right side). The contraction path follows the typical convolutional neural network structure, repeatedly performing two 3x3 convolutions operations (unpadded convolutions), accompanied by a rectified linear unit (ReLU) and a 2*2 max pooling operation with a stride of 2 called downsampling, in which the number of feature channels is doubled in time. Usually, the contracting path is replaced by a pre-trained network like VGG/Resnet.  In this work, our U-net architecture is composed of the Squeeze Excitation ResNet backbone, followed by an expansive path (see Figure \ref{unet}).

Squeez and Excitation ResNet network uses the concept of residual representation, commonly used in computer vision, to the construction of a basic block of residual learning in a CNN model. It uses some reference layers to learn the residual representation between input and output, rather than using a reference layer to directly learn the mapping between inputs and outputs as in a typical CNN network. Experiments have shown that using a reference layer to learn residuals directly is much easier and more efficient than learning input and output mapping, with a faster convergence speed and higher accuracy.

\subsubsection{Overlapped region detection}

The output probability map from  SER-Unet requires post-processing for overlapped nuclei separation. Thereby, an adaptive threshold is applied to binarize the probability map image. The adaptive method is an iterative algorithm that measures a local threshold determined by the mean grayscale intensities of a neighborhood. This method was preferred over other popular methods such as Otsu's method (Otsu, 1979) because it is local and can adapt to different luminance proportions. Next, the small holes in the binary image are filled, and a binary opening is performed to construct nuclei's shape. Finally, all small objects are removed. 

\begin{figure}[h]
\centering
        \includegraphics[totalheight =10cm]{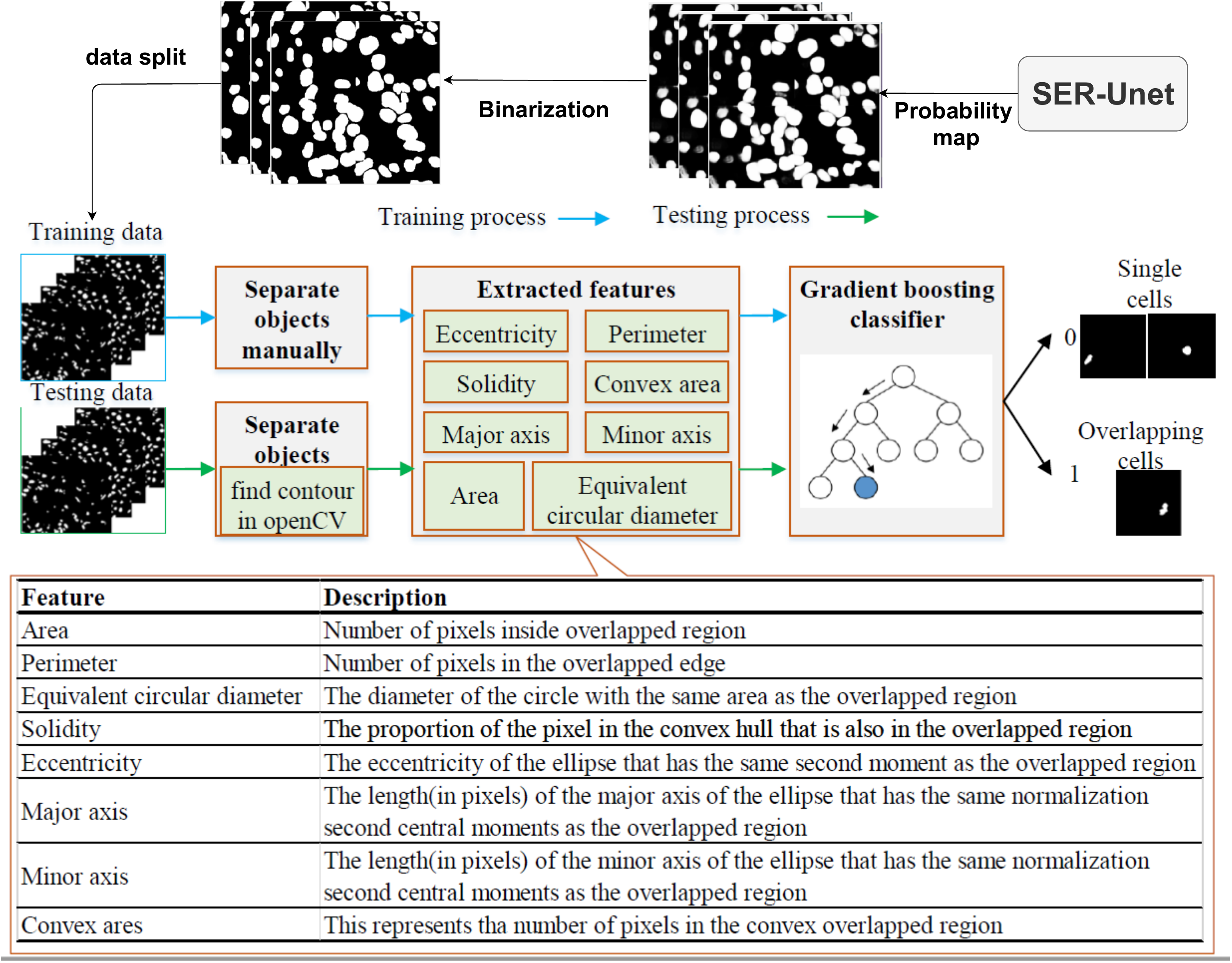}
    \caption{{The procedure of detection regions of overlapped nuclei using GB classifier.}}
    \label{fig:GDB}
\end{figure}

In our task, the generated probability map from the proposed algorithm depends on the density of the DeoxyriboNucleic Acid DNA in the nucleus. If the DNA is condensed in one part of the nucleus, the nucleus's mask takes shape with concavity. Thus, more specific separation based on concave points and markers based on h value \citep{koyuncu2016iterative} may lead to over-segmentation. In this case, we propose to apply separation techniques just on the overlapped regions to decrease the computational time and give good accuracy in our approach (see Algorithm \ref{algo}). In order to distinguish overlapped nuclei from single nuclei, features like size and shape are not sufficient. Therefore, we use textural and geometric features to train a binary classifier. The  SER-Unet model is used to predict masks from our train dataset. For each mask, we extract and classify the object manually into overlapped and single patches. Features are extracted from each patch to construct the dataset needed to train a classifier model (Figure \ref{fig:GDB}).

Gradient boosting (GB) \citep{friedman2002stochastic}  is a machine learning technique for classification problems, which produces a robust classifier in the form of an ensemble of weak ones, typically decision trees. GB builds the model in a stage-wise fashion, and it is a generalization to arbitrary differentiable loss functions. The GB classifier's performance in the task of overlapped region classification reached 92,69 \%, outperforms other algorithms such as support vector machines and RF.

\begin{algorithm*}[h]
{Input: Processed mask
Output: Mask with separated overlapped objects
Objects $\leftarrow$  get objects from processed mask; 
\For{object in objects}{
  
  \If{contours(object)\textgreater 10}{
        features $\leftarrow$ get features(object); 
        
        probability $\leftarrow$ gradient\_boost.predict(features);
        
  \eIf{probability\textless{}0.5}{
  
       single $\leftarrow$ draw(single, object);
       
       }{
       
       split $\leftarrow$ Marker\_based\_watershed(object);
       
       separated $\leftarrow$ draw(separated, split);
       
      }
   }
 }
 all-separated $\leftarrow$ merge(single, separated);
 
 return all-separated ;}
 \caption{Proposed separation approach}
 \label{algo}
 \end{algorithm*}

\subsubsection{Overlapped nuclei separation}

\begin{figure}[h]
\centering
        \includegraphics[totalheight = 9 cm]{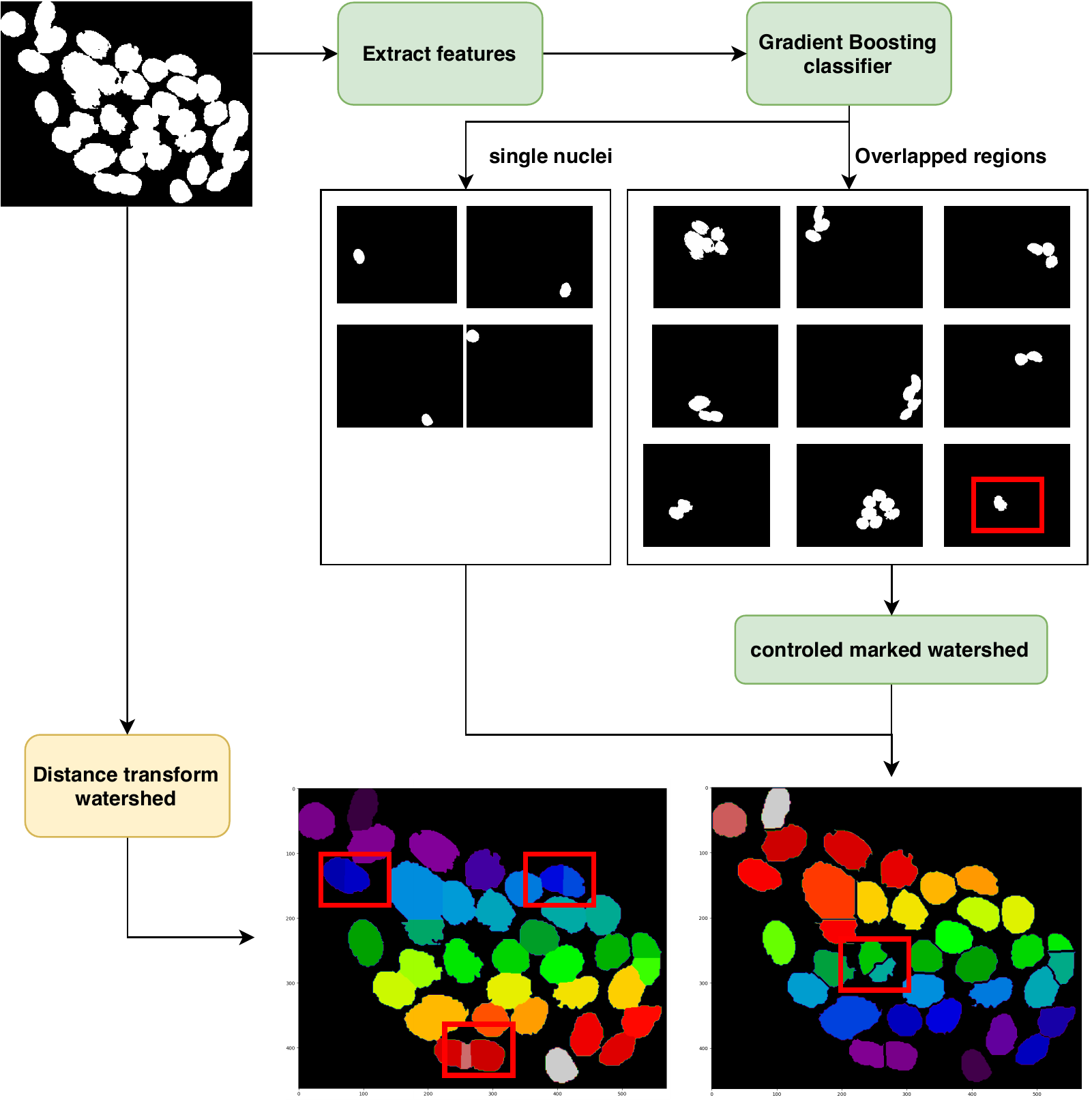}
    \caption{The proposed approach for overlapped nuclei separation. Gradient Boosting classifier is used to separate overlapped regions. Then, a marker-based watershed using iterative seeds is applied to separate overlapped nuclei.}
    \label{sep_stepss}
\end{figure}

Once the overlapped regions are separated, a watershed algorithm is used for nuclei split. Watershed transform algorithm is simple, reliable, and can reach acceptable results by splitting overlapped regions based on the image topology. However, the watershed transform is highly sensitive to local minima, and this leads to over-segmentation of the noisy image, particularly for medical data \citep{nagesh2010improved}. In this paper, we propose a simple iterative erosion with an elliptic kernel for markers selection. The problem of vanishing object in iterative erosion is solved by calculating the number of blobs after each erosion step; if the number of blobs is equal or higher than the previous one, we do another erosion step; if not, we get the vanishing blobs from the previous eroded image by copying the minimum of blobs based on the area in the seed image. The process is repeated until there is no blob in the eroded image. Finally, we dilate the final seed image and apply the transform watershed based on these markers (Algorithm \ref{algo}). Usually, successive erosion for watershed markers needs some stopping criteria. However, in our case, the stopping criteria is not required due to the small number of nuclei (between 2 and 5) in each overlapped region. Figure \ref{sep_stepss} presents the different steps of our proposed method on overlapped region \citep{kowal2020cell}. It is worth mentioning that we have different degrees of overlapping, and highly overlapped nuclei are difficult to be separated. However, in our dataset, this kind of overlapping is less frequent, and therefore it has a low impact on the PI scoring.



\subsubsection{Nuclei classification}

\begin{figure}[h]
\centering
        \includegraphics[totalheight = 2.5 cm]{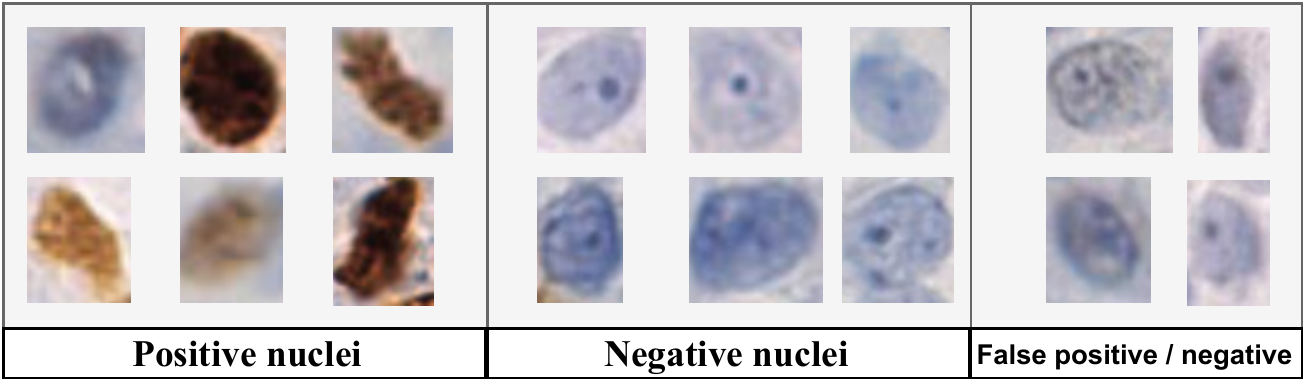}
    \caption{Nuclei variations generates high level of false positive and false negative cases.}
    \label{nuc_var}
\end{figure}

Image classification based on visual content, especially histopathologic images, is challenging due to complex textures and structural-morphological diversity. In H-DAB stained images, the classification of immunopositive and immunonegative cells is exposed to inter and intra-variability ({Figure \ref{nuc_var}}). Therefore, automating this process could reduce the effort and assist the expert to be more precise.  Recently,  CNN based approaches have achieved success in image classification problems, including medical image analysis \citep{bayramoglu2016deep}. CNNs explore the possibility of learning features directly from input data, avoiding hand-crafted features \citep{bengio2013representation}. 

Different CNN architectures are empirically applied to observe variation in model characteristics (network depth, layer properties, training parameters, etc.) for nuclei classification in our dataset.  In general, we saw that structures with residuals networks perform better than others. Residual networks avoid the vanishing gradients problem in deep networks, which help train these models on small datasets like in our case. Also, we observed that small networks perform better in our dataset, and Resnet18 gives better results than Resnet50 or other well-known models. However, evaluating the Resnet18 model with Kfold cross-validation shows high variance. Therefore, this study proposes a hybrid classifier by getting an advantage from features extracted from Resnt18 and avoiding the overfitting problem. 

Deep features are extracted using the Resnet18 classifier by removing the last dense layer. Then, a random forest (RF) classifier is used for nuclei classification. RF is a classification algorithm developed by Leo Breiman using a set of classification trees \citep{breiman2001random}. RFuses both bagging (bootstrap aggregation), a successful approach for combining unstable learners, and the selection of random variables for tree construction. As a result, the algorithm produces a set that can achieve both low bias and low variance. 

\begin{figure}[h]
\centering
        \includegraphics[totalheight = 7cm]{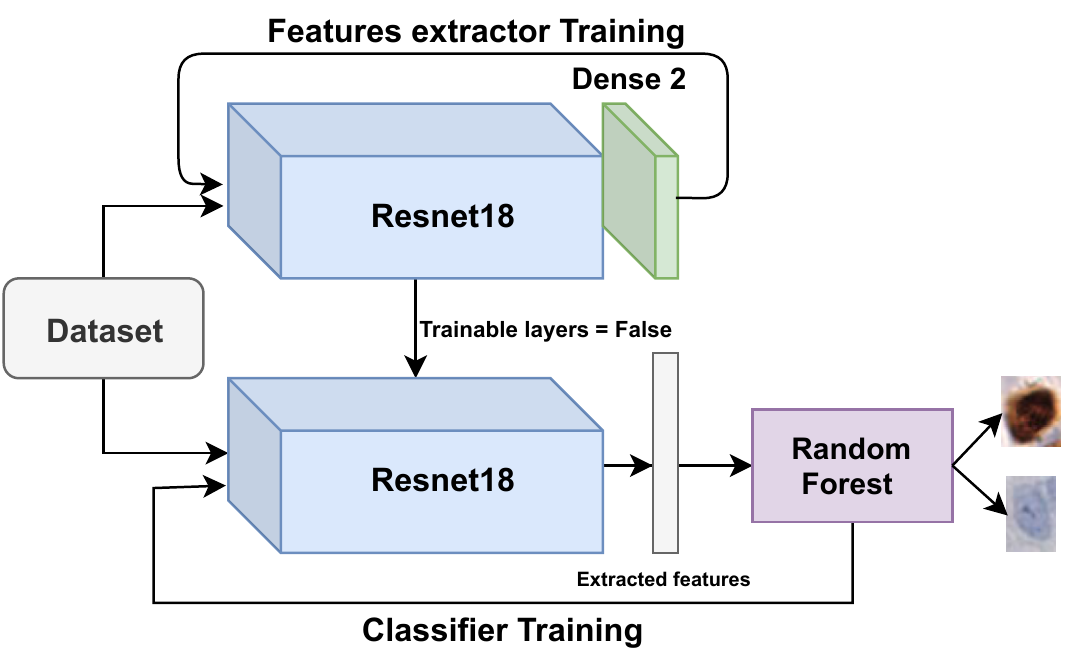}
    \caption{Proposed approach nuclei classification. Deep features are exstracted by training Resnet18 using transfer learning with Imagenet weights on a training set of multiple nuclei patches. A RFclassifier is then used to associate each patch into a positive or negative class. }
    \label{classifier}
\end{figure}

{CNNs with pre-trained weights have been used as feature extractors in previous works.}, and the classifier is then trained on these features \citep{gupta2020analysis}. In our approach, the training is two folds. First, the Resnet18 is trained with two classes on the dense top layer. Then, the last layer is removed, and the rest of the layers are set to trainable "False." Finally, the RF algorithm is trained using Resnet18 deep features (see Figure \ref{classifier}).

\subsubsection{Proliferation index scoring}

Cell proliferation is a fundamental biological process, and it is accelerated in tumors. Physicians use the proliferation index PI to assess tumor progression and determine a patient's future therapy \citep{hanahan2011hallmarks}. In this paper, the PI score is estimated by calculating the average percentage of positive tumor cells based on three regions:  Hotspot, medium, and edge, selected manually by an expert pathologist. The PI value is calculated using the following equation:

\begin{equation}
    PI =  \frac{1}{r}\sum_{i=1}^{r} \frac{N_{i}}{T_{i}} \times 100
    \label{equ:Ki-67}
\end{equation}
where $r$ is the number of selected regions of interest, $N_{i}$ is the number of immunopositive nuclei in region $i$ and  $T_{i}$ is the total number of nuclei in region $i$ {(see \cite[p. 1662]{dowsett2011assessment})}.

\subsection{Statistical analysis}

To validate the proposed methodology, three validation tests are conducted:

\subsubsection*{Segmentation metrics}

A good evaluation criterion for nuclear segmentation techniques should penalize both object-level and pixel-level errors \citep{komura2018machine}. Nuclei segmentation can pass through three main steps: 1) separate nucleus from the background, 2) detect individual nucleus, and 3) segment each detected instance. However, the evaluation of semantic results still an ongoing problem. In general, we can recognize, and use for comparison, two categories of evaluation metrics:

\subsubsection*{- Pixel level metrics:} 
The same metrics used as in \citep{long2015fully}, which are the accuracy (Acc), the mean intersection over union (MIU), and the frequency weighted intersection over union (FIU):

\begin{align}
Acc  &= \frac{\sum_{i} n_{ii}}{\sum_{i} t_{i}}\\
 MIU &= \frac{1}{n_{c}l} \sum_{i} \frac{n_{ii}}{t_{i} + \sum_{i} n_{ji} n_{ii}}\\
    FIU &= (\sum_{k}t_{k})^{-1} \sum_{i} \frac{t_{i} n_{ii}}{t_{i} + \sum_{i} n_{ji} n_{ii}}
\end{align}
where $n_{c}l$ : number of classes included in ground truth segmentation,  $n_{ij}$ : number of pixels of class i predicted to belong to class j and $t_{i}$ : total number of pixels of class i in ground truth segmentation.

\subsubsection*{- Object-level metrics:} 

For nuclei segmentation, a true positive TP is considered when an overlap of predicted and ground truth nuclei equal to or bigger than 0.5.  Most studies used ensemble dice (Dice 2) \citep{vu2019methods}, and Agreggated Jaccard Index AJI (\citep{komura2018machine}) for object-level segmentation. Recently, a new metric called panoptic quality PQ is introduced in (\citep{kirillov2019panoptic}) to evaluate segmentation results. Moreover, the credibility of Dice2 and AJI metrics is discussed, especially in the cell detection process. In (\citep{graham2019hover}), the author showed two similar cases with a slight difference that can give highly different scores of Dice2 and AJI. For the same case, panoptic quality gives a small deviation. Thereby, in this study, results are evaluated with Dice2, AJI, and PQ. 

PQ is calculated for each class independently and average over classes. This makes PQ insensitive to class imbalance. For each class, with a set of matching split composed of true positives (TP), false positives (FP), and false negatives (FN), PQ is defined as follows:

\begin{equation}
    PQ = \frac{\sum_{(p,g)\in TP} IoU(p,g)}{TP+ \frac{1}{2} FP +  \frac{1}{2} FN}.
\end{equation}

Furthermore, if PQ is multiplied and divided by the size of the TP set, then PQ can be viewed as the multiplication of a segmentation quality (SQ) term and a detection quality (DQ) term: 
\begin{equation}
    PQ = \frac{\sum_{(p,g)\in TP} IoU(p,g)}{TP} \times \frac{TP}{TP+ \frac{1}{2} FP +  \frac{1}{2} FN}.
\end{equation}
Addressed this way, the first term in the multiplication is segmentation quality (SQ), which is simply the average Intersection over Union IoU of matched segments; the second one is detection quality (DQ) or merely the well-known F1 score (\citep{van1986non}) broadly used for quality estimation in detection settings (\citep{martin2004learning}).

\subsubsection*{Classification metrics}

Given the number of true positives (TP), false positives (FP), true negatives (TN), and false negatives (FN), the performance of the proposed classification model evaluated based on accuracy, sensitivity, specificity, and F1-score mathematically expressed as follows:
\[
\systeme*{
    Accuracy = \frac{TP + TN}{TP+TN+FP+FN} \times 100,
    Sensitivity = \frac{TP }{TP+FN} \times 100,
    Specificity = \frac{TN }{TN+FP} \times 100,
    \text{F1-score} = \frac{2TP }{2TP+FP+FN} \times 100}
\]

\subsubsection*{PI metrics}

{PI scoring is performed using different methods, color based such as IHCCH, or machine leraning based such as Quapth, or deep learning based such as our proposed approach. The impact of overlapped nuclei separation is evaluated on these techniques based on the well-known Bland-Altman plots, Pearson, Spearman, R2 coefficients metrics for PI score estimation \citep{geread2019ihc}.}

\section{Experimental Results}
\label{results}
 
{\subsection{Study setup}
All algorithms are developed in python with the Keras library and the TensorFlow backend and are trained with a free Tesla K80 GPU available in Google Colab. For segmentation and classification models' training, the associated dataset has been divided according to patients into train, validation, and test. Five folds cross-validation was used, and the evaluation of the results was averaged from the test dataset.}

{For segmentation, SER-Unet algorithm is developed with transfer learning using Imagenet pre-trained weights. We defined the input image size of 256× 256 pixels. The training process has been done with 500 epochs in total, under the mini-augmented training batches with 250 train steps and 50 validation steps per epoch. We have used binary cross-entropy dice loss and Adam optimizer with learning rate decay. To avoid overfitting, early stopping with 20 epochs and save the best models strategies have been used. Finally, the model is saved and used for test evaluation. Similar to segmentations evaluation, the same ten patients are used as a test set for classification. From the three images of each patient, nuclei are cropped based on expert annotation. Models are trained successively in a cross-validation mode on the training set, and evaluation is then performed on the test set.}

\subsection{Segmentation results}

{Table \ref{sem_res} presents the first evaluation of the segmentation results against recent deep neural networks, mainly LinkNet \citep{chaurasia2017linknet} and PSPNet \citep{zhao2017pyramid}.} 

\begin{table}[h]
\centering
\caption{semantic segmentation results}
{\begin{tabular}{lllllll}
\hline
                 & DICE2 & AJI   & PQ    & Acc   & MIU   & FIU   \\ \hline
Unet             & 68.72 & 55.14 & 51.31 & 90.52 & 81.43 & 81.21 \\ \hline
LinkNet          & 67.95 & 51.69 & 45.67 & 90.88 & 79.07 & 83.45 \\ \hline
PSPNEt           & 68.08 & 54.16 & 48.73 & 91.22 & 79.18 & 83.71 \\ \hline
SER-Unet & 71.86 & 59.21 & 55.42 & 92.46 & 84.15 & 85.13 \\ \hline
\end{tabular}}
\label{sem_res}
\end{table}

\begin{figure}[h]
\centering
        \includegraphics[totalheight =3.4cm]{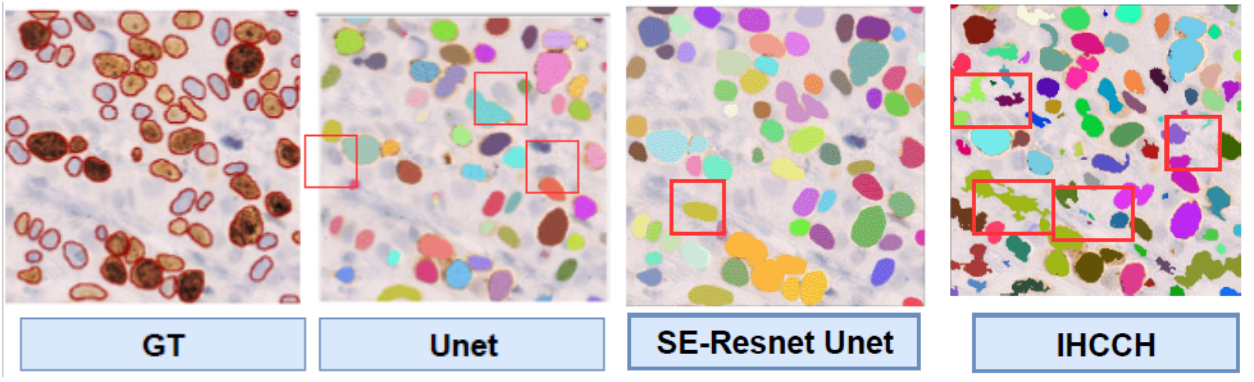}
    \caption{Semantic segmentation results: good detection of fully convolutional networks, especially with the proposed approach combining squeez excitation residual network with Unet algorithm.Conventional methods suffer from oversegmentation due to the poor quality of staining, and the presence of infalammation cells with same color as negative nuclei.}
    \label{sem_seg}
\end{figure}

{Figure \ref{sem_seg} shows the difference between the proposed approach and conventional techniques segmentation \citep{geread2019ihc}, which are highly sensitive to color space. Inflammation nuclei with blue color are considered negative nuclei due to their color similarity. However, our proposed approach that combines Squeez Excitation Residual networks with Unet algorithm is robust and can discriminate correctly negative and positive nuclei based on colors and the associated shape learned from annotated data. Hence, it gives better results, especially for object-level metrics, with a PQ equal to 55.42\%. }

\subsection{Overlapped nuclei separation results}

{Most of the overlapped nuclei are not separated in the probability map generated with our proposed approach, and more post-processing is required. Accordingly, we apply and compare Gaussian Mixture Model (GMM), Distance Transform Watershed (DTW), and our proposed marker-based watershed algorithm for overlapped nuclei split. Table \ref{prrec} shows the quantitative results of overlapped nuclei separation based on Precision, Recall, and F1-score.}
\begin{table}[h]
\caption{Segmentation results with  SER-Unet after separation of different techniques.}
\centering
{\begin{tabular}{lllllll}
\hline
         & DICE2   & AJI     & PQ      & Precision & Recall  & F1 score \\ \hline
GMM      & 38.44 & 32.09 & 10.21 & 35.64   & 30.66 & 32.97  \\ \hline
DTW      & 78.46& 61.54 & 57.39 & 78.29   & 63.96 & 70.40  \\ \hline
Proposed & 79.60 & 73.01 & 69.80 & 87.45   & 84.15& 85.77  \\ \hline
\end{tabular}}
\label{prrec}
\end{table}

\begin{figure*}[h]
\centering
        \includegraphics[totalheight = 13cm]{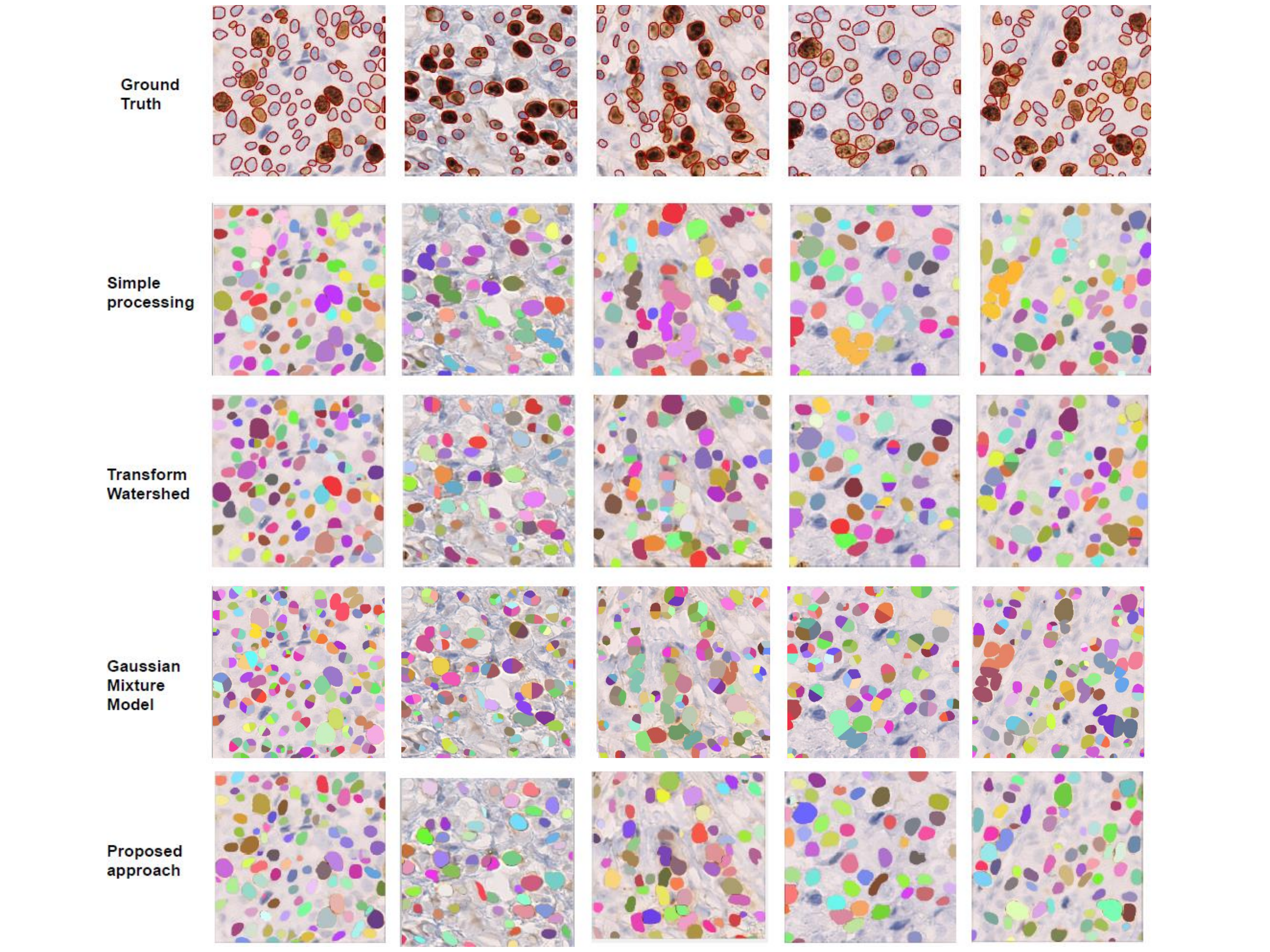}
    \caption{Comparison of the proposed approach of Overlapped nuclei separation with different techniques: Simple processing with simple filters give under-segmentation. On the other hand, Transform Watershed and GMM separate nuclei incorrectly including single nuclei. The proposed approach in the last row give better results.  }
    \label{vis}
\end{figure*}

Figure \ref{vis} shows the visual comparison between our algorithm, GMM, and DTW in terms of segmentation results.

\subsection{Classification results}

\begin{figure}[h]
\centering
        \includegraphics[totalheight = 5cm]{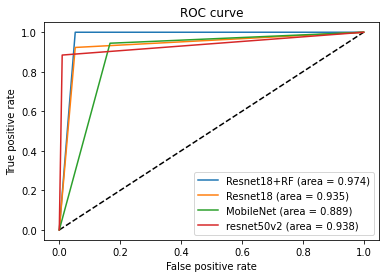}
    \caption{ROC curve and AUC of the proposed approach in comparison with different classifiers. }
    \label{roc}
\end{figure}

Table \ref{class_res} shows the performance of the proposed approach with recent deep learning classifiers.  The proposed hybrid approach of deep features extraction and RF classifier gives the best results with F1-score 97.52\%. Moreover,  Figure \ref{roc} shows the classifiers' ROC curve, which shows that the proposed approach also gives the highest AUC with 0.974, against Resnet18 with 0.935 and MobileNet with 0.889.  

\begin{table}[h]
\centering
\caption{Classification results}
{\begin{tabular}{lllll}
\hline
                & F-score & Accuracy & Sensitivity & Specificity \\ \hline
MobileNet       & 89.13   & 89.19    & 83.33       & 94.44       \\ \hline
InceptionResnet & 92.56   & 92.57    & 97.14       & 88.46       \\ \hline
Resnet50        & 93.46   & 93.47    & 99.05       & 88.46       \\ \hline
Resnet18        & 93.47   & 93.47    & 94.76       & 92.31       \\ \hline
Resnet18+RF     & \textbf{97.51}   & \textbf{97.52 }   & \textbf{94.76 }      & \textbf{100 }      \\ \hline
\label{class_res}

\end{tabular}}
\end{table}

\subsection{PI results}

\begin{figure}[h]
\centering
        \includegraphics[totalheight = 20cm]{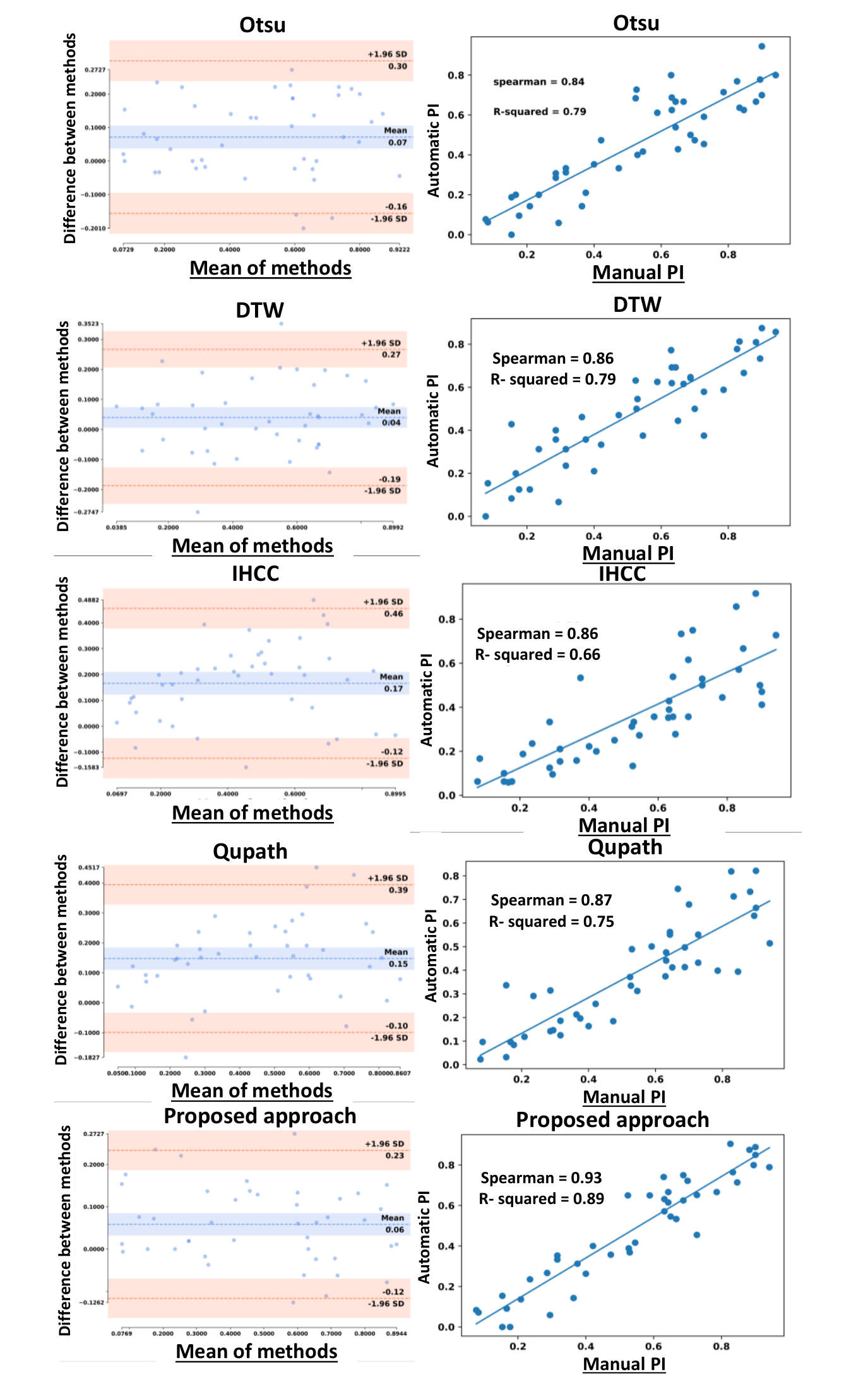}
    \caption{Spearman, R-square, and Bland-Altman results of the predicted PI and and manual PI for different methods. The difference between methods for Bland-Altman plots stands for (manual PI - automatic PI). Strong relationship between the predicted PI with the proposed approach and manual PI.}
    \label{scatter}
\end{figure}

{Figure \ref{scatter} illustrates the PI agreement's scatter plots between the manual and automated approaches on the test set images. R2, Spearman, and Pearson correlations were computed for each approach. To show the importance of post-processing on PI estimation, the otsu and DTW post-processing methods are compared with the proposed approach, in which the GMM method is excluded due to its poor performance. Also, the proposed histogram-based method (IHCCH) in \citep{geread2019ihc} is used for further analysis. The IHCCH is an unsupervised color separation method based on the IHC color histogram for the robust analysis of PI.}

{Figure \ref{scatter} shows that the proposed deep learning approach gives better results than the color-based approach, even with simple post-processing like the Otsu method. The proposed approach obtained the highest Spearman correlation coefficient with 0.93; and R2 = 0.89. Hence, it indicates a strong relationship between the predicted PI and the manually counted PI.}

The corresponding Bland-Altman plots were calculated based on the difference in PI between the automated and manual approaches and are shown in Figure \ref{scatter}. The optimal threshold is the one with the lowest average PI difference along the narrow boundaries of the agreements. Although the boundary has relatively small PI differences, the proposed approach gives the best results.

\begin{figure*}[t]
\centering
        \includegraphics[totalheight = 20cm]{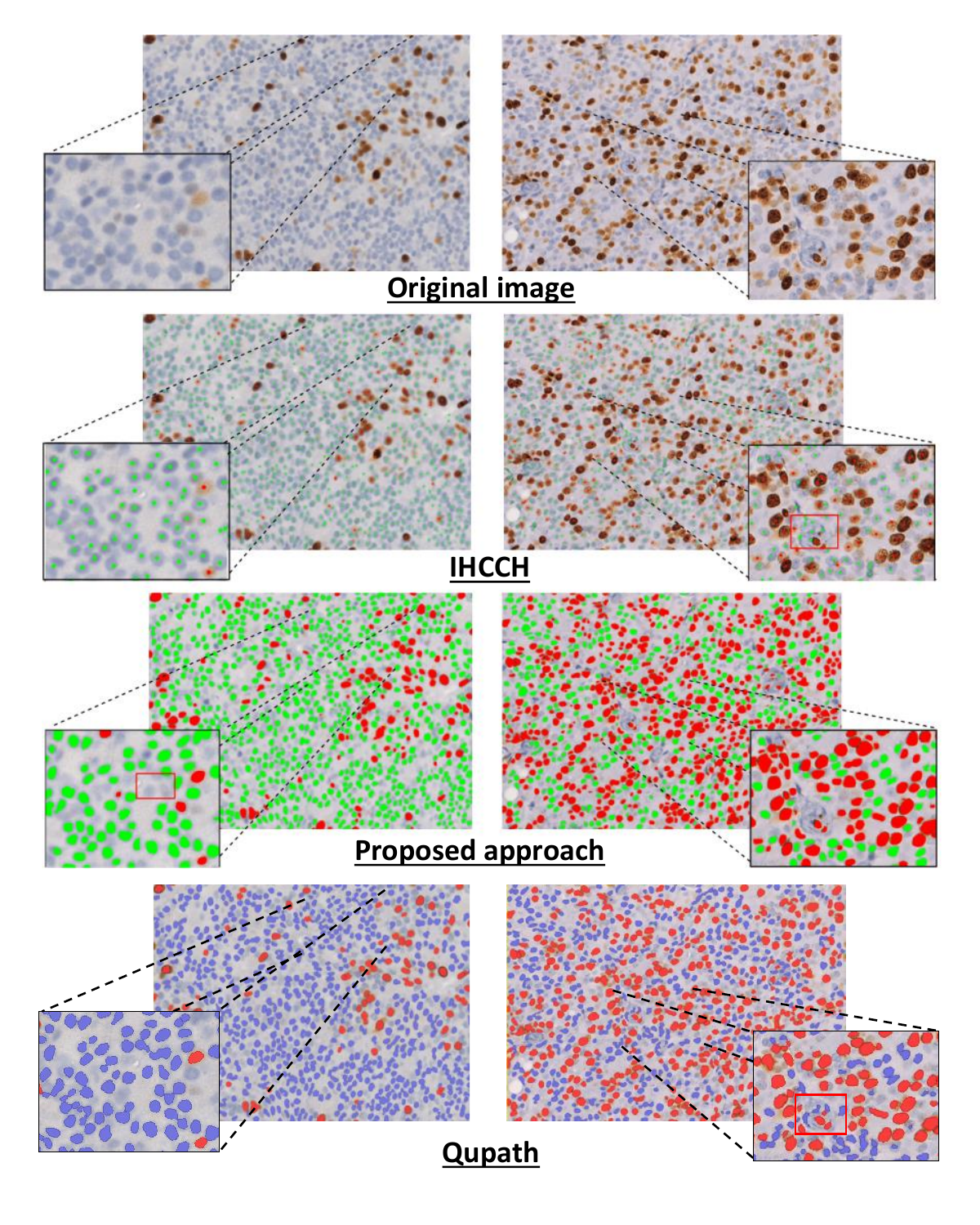}
    \caption{Qualitative evaluation of PI estimation. (First row): ground truth of two regions with different proliferation rate. (Second row): IHCCH results. (Third row): Proposed approach results. (Fourth row): Qupath results. The IHCCH method tends to over-segment compared to the proposed approach, which is robust towards non nuclei regions with nuclei-like color.} 
    \label{quali}
\end{figure*}

{For qualitative assessment, the original image, the IHCCH method, Qupath tool, and our proposed method findings are presented in Figure \ref{quali}.  Immunopositive and immunonegative results are shown in red and green, respectively. The findings suggest that our approach segmentations are visually similar and produce natural boundaries. }

\section{Discussion}
\label{disc}

Computer-assisted pathology has increasingly shown its advantages over the years. For early detection of breast cancer, counting cells or nuclei is one of the most used techniques. However, the automatic nuclei counting in histopathologic images faces challenges such as inter- or intra-nuclei color variations, irregular shapes, and touching or overlapping nuclei. We here proposed a fully automated PI estimation pipeline to analyze histology images of breast cancer. Note that the most critical cases for the experiment phase were with images with overlapped nuclei. In such a situation, separation techniques might not ensure accurate split, especially where the predicted map probability of one nucleus takes a concave shape and is considered two nuclei. Results show that the problem of overlapping nuclei can affect the performance of PI estimation even with deep learning techniques for detection. 

{Initial results confirm that FCNs are a good choice for medical image segmentation. Their main advantage is the segmentation based on different features such as shape, stain, and other deep features extracted by successive convolutions. For separation task, Table  \ref{prrec} shows that GMM has the worst results due to over-segmentation. Also, DTW yielded a relatively lower performance than the proposed approach. This is mainly due to difficulties incurred by nuclei's random morphology, making it challenging to find the right split. Nevertheless, our proposed algorithm gives the best results, especially in terms of PQ. Moreover, the post-processing of SER-Unet with the proposed approach increases the model's performance significantly, especially for Dice2 and AJI, which are too sensitive to the small variation.}

{Color-based methods are affected by many low-level processing elements, including stain intensity thresholds and nuclei segmentation accuracy. As a result, most of the predicted PIs by the IHCCH method are lower than the ground truth PI, which can be explained by the over-segmentation of negative nuclei. On the other hand, Qupath applies separation techniques on the segmented nuclei using conventional methods and reaches 0.75 of R2 and 0.87 of Pearson coefficient, which is lower than the combination of separation techniques with deep learning solutions. The proposed approach obtained the highest Spearman correlation coefficient with 0.93 and R2 = 0.89. Hence, it indicates a strong relationship between the predicted PI and the manually counted PI. On the other hand, our approach is sensitive in cases of weak staining. As shown in Figure \ref{quali}, some weakly colored nuclei lose their shape features, which is quite essential in our approach to build deep features; therefore, nuclei are missed and considered as background.}

Moreover, histology image analysis faces numerous challenges. Stain variation is one problem we encounter when different pathology laboratories stain tissue slides that show similar but not identical color appearance. Due to this color variation between laboratories, CNNs trained with images from one laboratory usually under-perform on unseen images from the other laboratory. Many methods have been proposed to decrease the generalization error, mainly classified into two categories: stain color normalization and stain color augmentation. The former aims to suit training and test color distributions to overcome the stain variation problem. The latter reproduces a wide variety of realistic stain variations during training, producing stain-invariant CNNs.

In this work, we quantified these techniques' effect on the performance of the two CNNs models used for segmentation and classification. The two models are trained with/without normalization, including simple or strong augmentation. Results showed that simple augmentation without normalization is enough for the classification task. For the segmentation task, PQ of simple augmentation with normalization is 59.12  and without normalization is 51.31. As well, PQ of strong augmentation with normalization is 69.8 and without normalization is 69.2. Accordingly, we can notice that the SER-Unet model presented a high sensitivity to the stain and morphology variation. Thus, applying strong augmentation helped the model generalize and perform well on new unseen images. However, image normalization in this dataset did not enhance the model performance when using strong augmentation with it.

The presented study shows the impact of overlapped nuclei on PI scoring of breast cancer. However, in our framework, regions of interest are selected manually by an expert pathologist. {Furthermore, the presence of overlapping nuclei is not only dependent on the type of regions but also on the early stages of tissue preprocessing, so that the number of overlapping nuclei increases with increasing tissue thickness.} Future studies will explore the relationship between the PI scoring and other previous steps in the process. Also, we will focus on the automation of regions of interest detection and investigate the relationship with overlapped areas to accurately estimate the PI score.

\section{Conclusion}

In this paper a digital imaging approach is proposed to support medical experts and and reduce inter-observer and intra-observer variabilities. The fully automated pipeline estimates the PI to assist pathologists in their decisions. The proposed approach was validated on a dataset from the department of pathology of HNFC hospital in France and achieved the best results with Spearman's score equal to 0.93 and R2 equal to 0.89.

While the proposed approach presented in this work have demonstrated promising results, the proposed method cannot be applied directly to the whole slide image and necessitates a manual selection of regions of interest. Future research might extend the proposed approach to the automatic selection of regions of interest. Furthermore, future work will focus on the evaluation of the proposed approach on different slides from different laboratories. Finally, while deep learning approaches give good performance in DP, labeled data is required to train models; and we should mention that a significant factor in the development of more reliable and robust automated methods of DP image analysis will be accelerated through the engagement of the community of expert pathologists in the process of deep learning models development.

\section*{Acknowledgements}
This work has been supported by the Profas B+ scholarship and EIPHI Graduate school (contract "ANR-17-EURE-0002"). The authors would like to thank Etienne Martin, a biomedical engineer, for his valuable help.

\bibliographystyle{unsrt}
\bibliography{template}  






\end{document}